\documentclass[twoside]{dis08}
\usepackage[latin1]{inputenc}
\usepackage[dvips]{graphicx,epsfig,color}
\usepackage{wrapfig,rotating}
\usepackage{amssymb,amsmath,array}

\newcommand{\be}{\begin{equation}}
\newcommand{\ee}{\end{equation}}

\usepackage{cite}
\usepackage{amsmath}
\usepackage{graphicx}
\usepackage{epsfig}
\usepackage{hyperref}
\usepackage{url}
\usepackage{amssymb}

\newcommand{\Qa}[1]{$Q_{f=#1}^{w}(R)~$}

\newcommand{\bea}{\begin{eqnarray}}
\newcommand{\eea}{\end{eqnarray}}

\newcommand{\lp}{\left(}
\newcommand{\rp}{\right)}

\begin{document}
\title{Quantifying the performance of jet algorithms at LHC}

%
%
%
%
%
%

\author{Juan Rojo
%
%
\vspace{.3cm}\\
%
Laboratoire de Physique Theorique et Hautes Energies (L.P.T.H.E.) \\
UPMC - Paris VI, 4 place Jussieu, \\
F-75252 Paris Cedex 05 France 
}

\maketitle

\begin{abstract}
In the present contribution
we introduce a strategy to quantify the
performance of modern infrared and collinear safe jet clustering 
algorithms in processes which involve
the reconstruction of heavy object decays.
We determine optimal choices 
for fictional narrow $Z'\to q\bar{q}$ 
and $H\to gg$ over a range of
masses,
providing examples of simple quark-jet  and 
gluon-jet samples respectively.
We show also that our estimates are robust against the
presence of high-luminosity pileup.
\end{abstract}

\paragraph{Introduction}
There has been sizable progress in jet algorithms in the recent
years \cite{Cacciari:2005hq,Cacciari:2007fd,Salam:2007xv,
Cacciari:2008gn,Cacciari:2008gp,Ellis:2007ib}. 
However, less work has been devoted with modern tools
to  determine the optimal jet algorithm (and 
associated parameters like $R$) for different physics processes
at the LHC.  
This contribution reports
on ongoing studies\footnote{Initial results have been presented in the
Les Houches 2007 workshop proceedings \cite{Buttar:2008jx}.}, in collaboration
with M.~Cacciari, G.~Soyez and G.~Salam, to
quantify the performance of modern jet algorithms and related
background subtraction strategies in the LHC environment
 in the case in which masses of heavy
particles are being reconstructed \cite{jetalgs}. 

\paragraph{General strategy}

We should recall that 
when studying the performance of jet algorithms, 
one should avoid figures of merit based on ambiguous concepts
like parton momenta and direction
(ill-defined in pQCD) or which assume a given distribution 
for the reconstructed mass spectra (like a Gaussian shape).
Instead, we shall use figures of merit related to
the maximisation of the signal over background ratio (more precisely,
$S/\sqrt{B}$).


The first  figure of merit is denoted by {\bf $Q_{f=z}^{w}(R)$}, the 
width of the smallest mass window that
  contains a fraction $f=z$ of the generated massive objects,
\be
\label{eq:q1}
f = 
 \lp \frac{{\rm \#~reconstructed.~massive~objects~in~window~of~width~}w}{
\rm Total ~\#~generated~massive~objects}\rp \ .
\ee
A jet definition that is more effective in reconstructing the majority
of massive objects within a narrow mass peak gives a lower value for
$Q_{f=z}^{w}(R)$, and is therefore a ``better'' definition.

The second figure of merit is denoted by
 {\bf $Q_{w=x\sqrt{M}}^{f}(R)$}. To compute this quality measure, 
we displace over the mass distribution a window  of fixed
width given by $w=x\sqrt{M}$, where $M$ is the nominal heavy object mass
that is being reconstructed, and we find 
the maximum number of events of the mass distribution contained in it.
In this situation we define this figure of merit as
\be
Q_{w=x\sqrt{M}}^{f}(R) \equiv
 \lp \frac{{\rm Max ~\#~reconstructed~massive~objects~in~window~of~width~} w=x\sqrt{M} }{
\rm Total ~\#~generated~massive~objects}\rp^{-1} \ .
\ee

To obtain a more physical interpretation,
the ratio of quality measures can be mapped to variations in effective
luminosity needed to achieve constant signal-over-background ratio
for the mass peak reconstruction. We assume that
 the background is flat and constant,
and unaffected by the
jet clustering. We define the effective power to discriminate
signal over background $\Sigma^{\rm eff}$ 
for a given jet definition (JA,$R$) as
$\Sigma^{\rm eff}\lp {\rm JA},R\rp \equiv N_{\rm signal}/
\sqrt{N_{\rm back}} $.
Then, for example in the case of \Qa{z}, if we define 
\be
r_w \equiv \frac{Q_{f=z}^{w}\lp{\rm JA_2}, R_2\rp}
     {Q_{f=z}^{w}\lp{\rm JA_1}, R_1 \rp}
=   \frac{N_{\rm back}\lp {\rm JA_2}, R_2 \rp}
        {N_{\rm back}\lp {\rm JA_1}, R_1 \rp} \ ,
\ee
at equal luminosity the discriminating 
power for $(\mathrm{JA}_1,R_1)$ will  
differ by a factor
$\Sigma^{\rm eff}\lp {\rm JA_1}, R_1 \rp/$
$\Sigma^{\rm eff}\lp {\rm JA_2}, R_2 \rp= \sqrt{r_w}$
with respect  $(\mathrm{JA}_2,R_2)$ . 
Equivalently the same discriminating power as $(\mathrm{JA}_2,R_2)$
can be obtained with a different luminosity ${\cal L}_1 = \rho_{\cal L} {\cal
  L}_2$, where $\rho_{\cal L} =  1/r_w$.

\paragraph{Jet algorithms}

We study the performance of  available IRC safe
jet algorithms: $k_T$ \cite{Catani:1991hj}, 
Cambridge/Aachen \cite{Dokshitzer:1997in,Wobisch:1998wt}, 
anti-$k_T$ \cite{Cacciari:2008gp} and SISCone \cite{Salam:2007xv}.
 On top
of these, we will examine also the performance of
the  filtering jet finding strategy, first introduced in
 \cite{Butterworth:2008iy}, with
$R_{\rm filt}=R/2$ and $n_{\rm sj}=2$ (labeled as C/A(filt) in
the various plots).

\paragraph{Processes investigated}
This general strategy has been
applied to both fictitious narrow $H\to gg$ and $Z'\to q\bar{q}$
decays, which provide examples
of physical gluon and quark jet samples
respectively. We consider 
a wide range of the heavy particle masses\footnote{Some of them already
excluded by measurements or indirect constraints, however in
the present work we use them as a source of mono-energetic jets only.}. 
 Also multijet events from hadronic $t\bar{t}$
 have been studied \cite{Buttar:2008jx}.

\begin{center}
\begin{figure}
\includegraphics[width=0.49\textwidth]{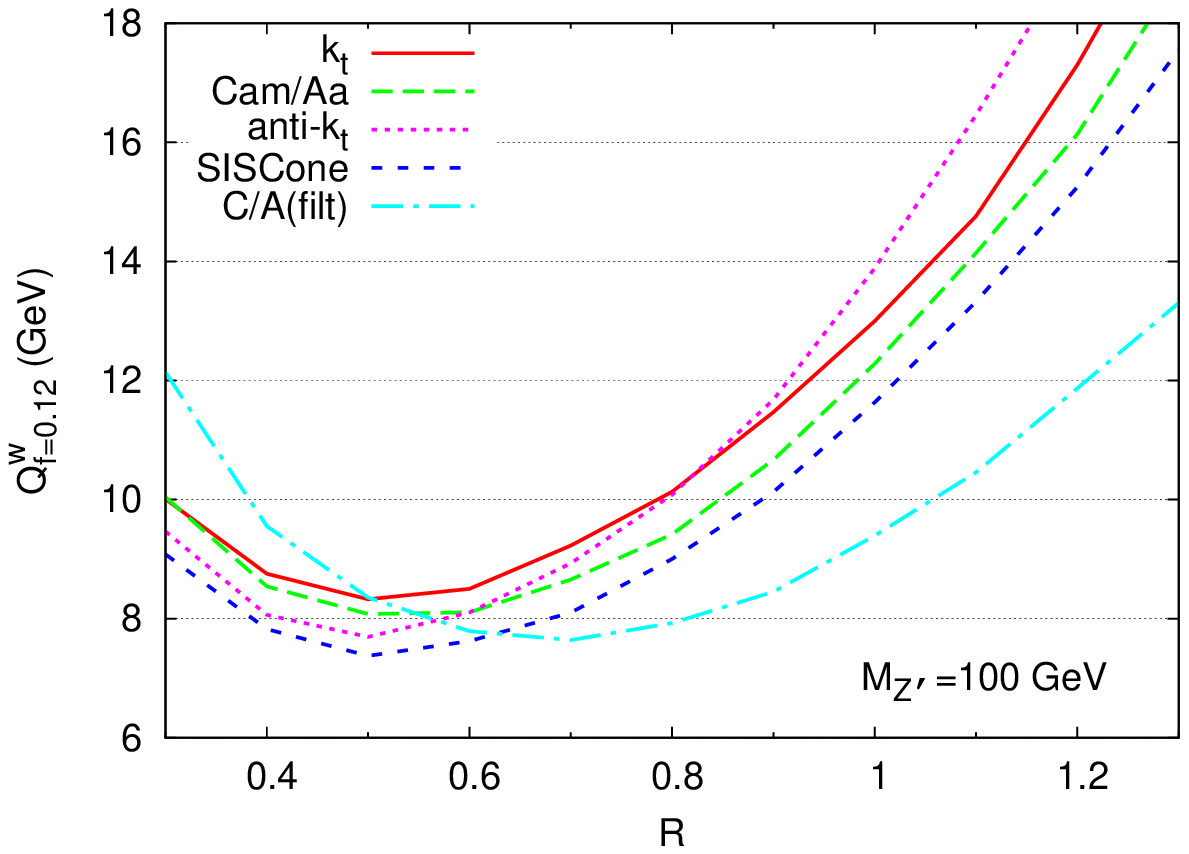}
\includegraphics[width=0.49\textwidth]{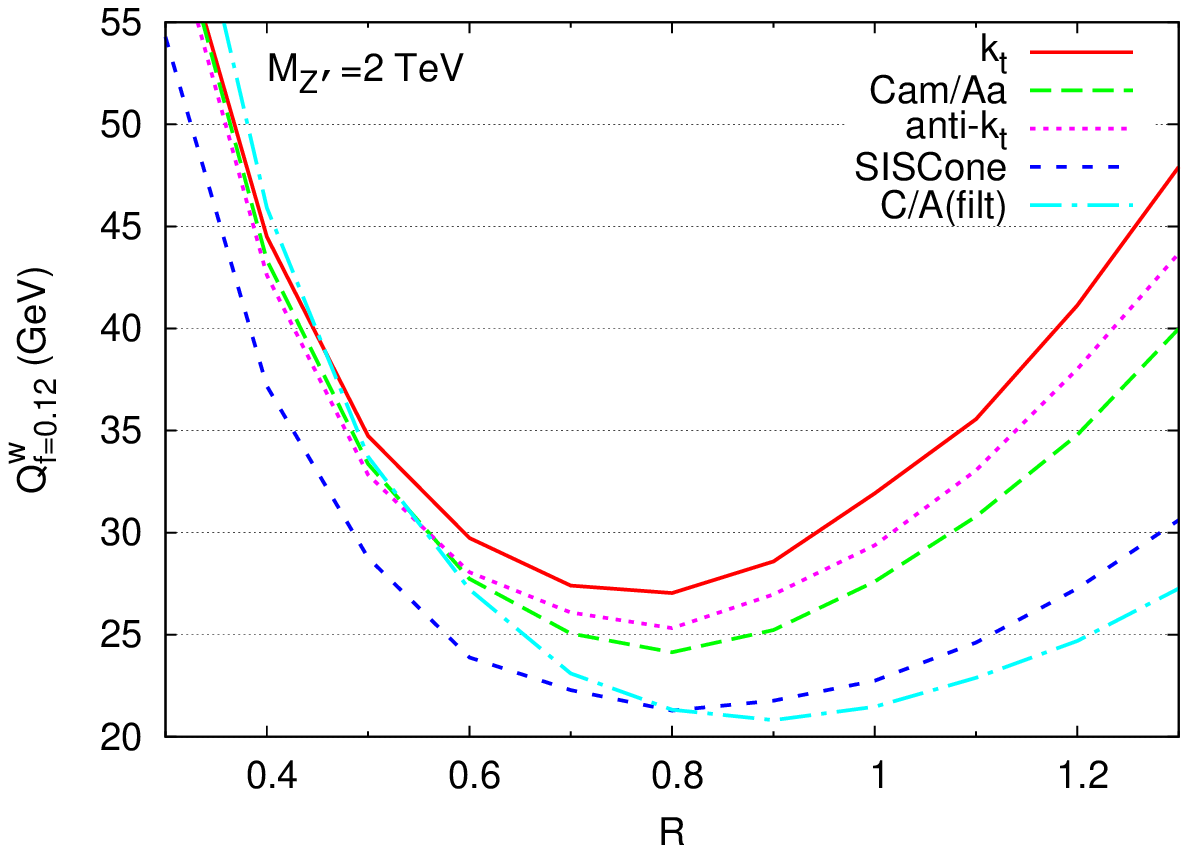}
\caption{\small  The figure of merit \Qa{z}
 for the quark jet samples from $Z'$
    corresponding to $M=100$~GeV (left plot) 
and $M=2$ TeV (right plot).\label{fig:quality_R} }
\end{figure}
\end{center} 

\begin{center}
\begin{figure}
\includegraphics[width=0.49\textwidth]{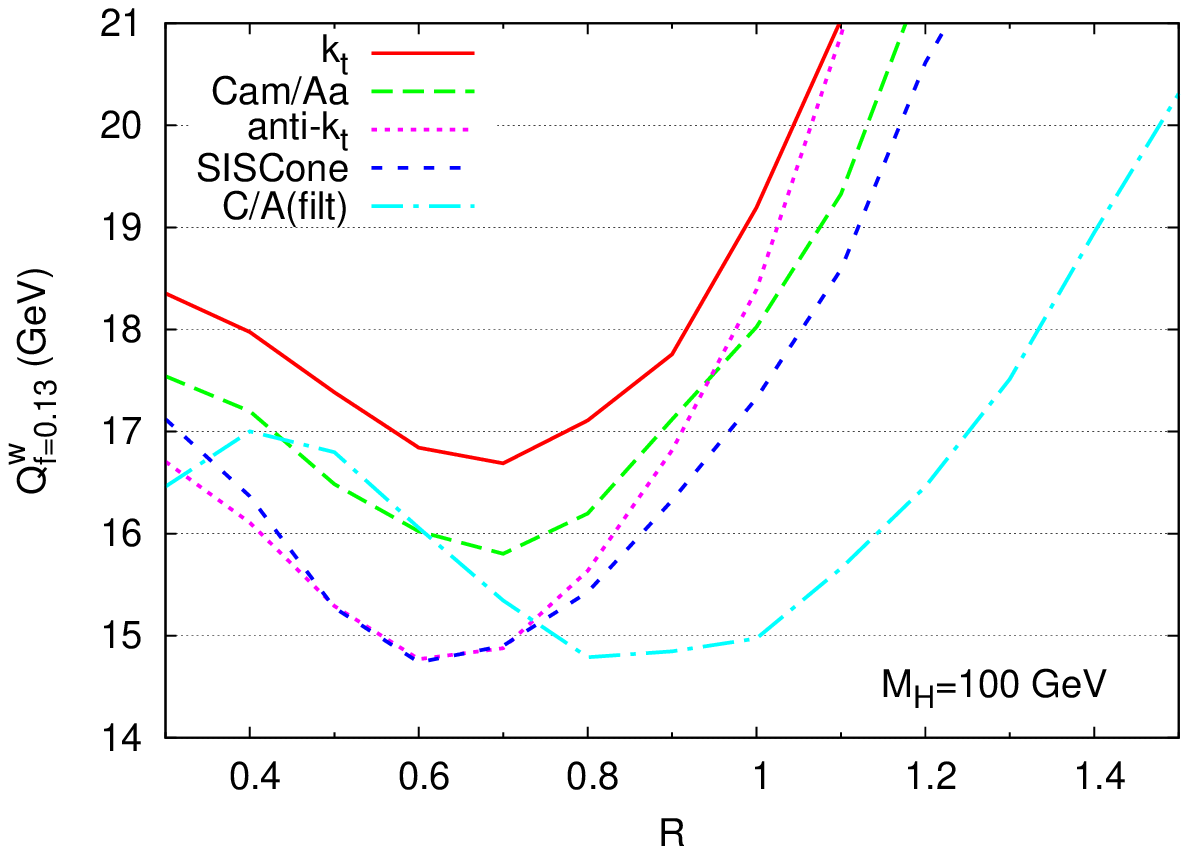}
\includegraphics[width=0.49\textwidth]{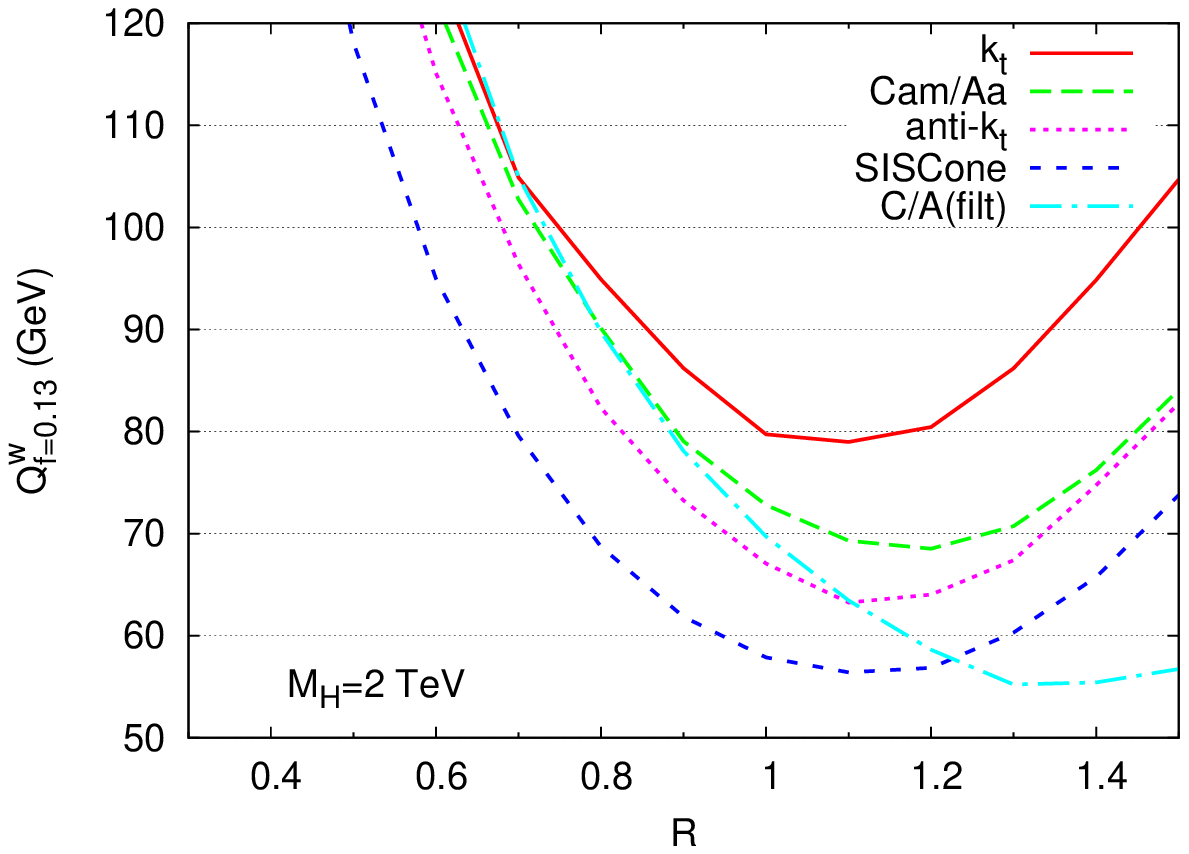}
\caption{\small  The figure of merit \Qa{z}
   for the gluon jet samples from $H$ 
    corresponding to $M=100$~GeV (left plot)
and  corresponding to $M=2$~TeV (right plot) 
\label{fig:quality_R2} }
\end{figure}
\end{center}

\paragraph{Results}
We show in Figs. \ref{fig:quality_R} and \ref{fig:quality_R2}
the quality measure \Qa{z} 
 for all five algorithms considered, both for quark
jets and gluon jets.
We observe in each case minima which define
the optimal value of the radius parameter $R_{\rm best}$. 
Note that the sources of quality difference do
stem either from the choice of jet algorithm 
(specially for gluon jets) as well as
from the value for $R$ adopted. Note that the results obtained
with the two quality measures are consistent.

In Fig. \ref{fig:RbestPU} 
we summarize the results for $R_{\rm best}$ for all jet algorithms
for gluon jets. We observe an approximately
scaling  $R_{\rm best}\sim \ln M_{H}$ ($p_T^{\rm jet}\sim M_H/2$), 
which can be understood due
to the contribution from
QCD perturbative radiation \cite{Dasgupta:2007wa}. The values found
satisfy  $R_{\rm best}\ge 0.7(0.9)$ for $p_T\ge 250$ GeV
quark(gluon) jets.

Let us examine less favored choices for the
jet definitions in the $M_{H}=2$ TeV case:
if we use SISCone, but with $R_{\rm best}^{100~\rm{GeV}}=0.6$
instead of $R_{\rm best}^{2~\rm{TeV}}=1.1$, we find 
 $\rho_{\mathcal{L}}\sim 0.55$. If on the other hand  we  
use $ R_{\rm best}^{2~\rm{TeV}}$,
but choose 
 {$k_T$} and instead of SISCone then
 $\rho_{\mathcal{L}}\sim 0.6$.
So we observe that almost half of the effective 
discriminating power $\Sigma^{\rm eff}$ is lost with these
choices.


We have studied as well how robust are our results
 with respect to the presence of Pile-Up (PU).
To this purpose, 
we generated minimum bias samples with Pythia Tune DWT for 
LHC at high luminosity, $\mathcal{L}_{\rm high}=0.25 {~\rm mb}^{-1}$
per bunch crossing.
PU is subtracted based on the jet area method \cite{Cacciari:2007fd}.
Our analysis shows that  even at high luminosity
the preferred values of $R$ are rather close to their
original values
without PU, as can be seen in Fig. \ref{fig:RbestPU}.

\paragraph{Summary}

We have presented a general strategy to quantify the performance
of jet algorithms in the case in which a heavy particle
mass is reconstructed.
We have shown that the  optimal jet definition, both
the jet algorithm and its  parameters like $R$, 
depend on 
both the kinematics of the process and the
mass scales involved. In the case of the dijet
samples studies, we find that
larger $M$ implies larger $R_{\rm best}$ to maintain
jet resolution. 

In our study {SISCone} and {C/A(filt)
turn out to be the optimal choices for these processes.
However,  these conclusions do in general depend
on the physics process considered, and can be
rather different in more complex multijet situations
like in hadronic $t\bar{t}$ production, as shown
in \cite{Buttar:2008jx}.
We have also checked that our quantitative estimates for
$R_{\rm best}$ are  robust  in the presence of high
luminosity PU after subtraction.

Let us finally emphasize again that these results have been obtained
with the assumption that the background is flat and
unaffected by jet clustering. Although our analysis cannot
in any case replace a proper $S/\sqrt{B}$ study, it 
is indicative
of the potential relevance of such variations in more
realistic studies, and emphasizes the importance of
flexibility for jet finding at the LHC.

\begin{center}
\begin{figure}
\includegraphics[width=0.49\textwidth]{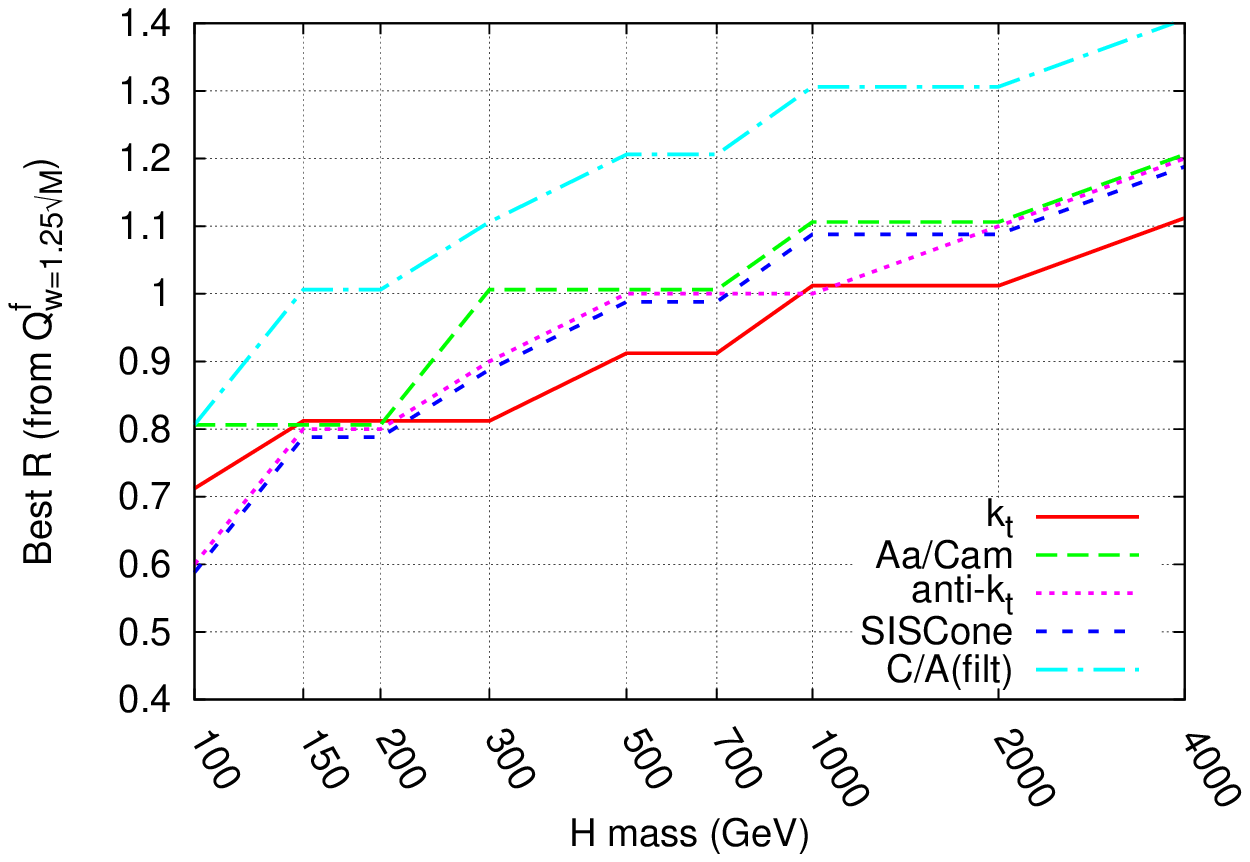}
\includegraphics[width=0.49\textwidth]{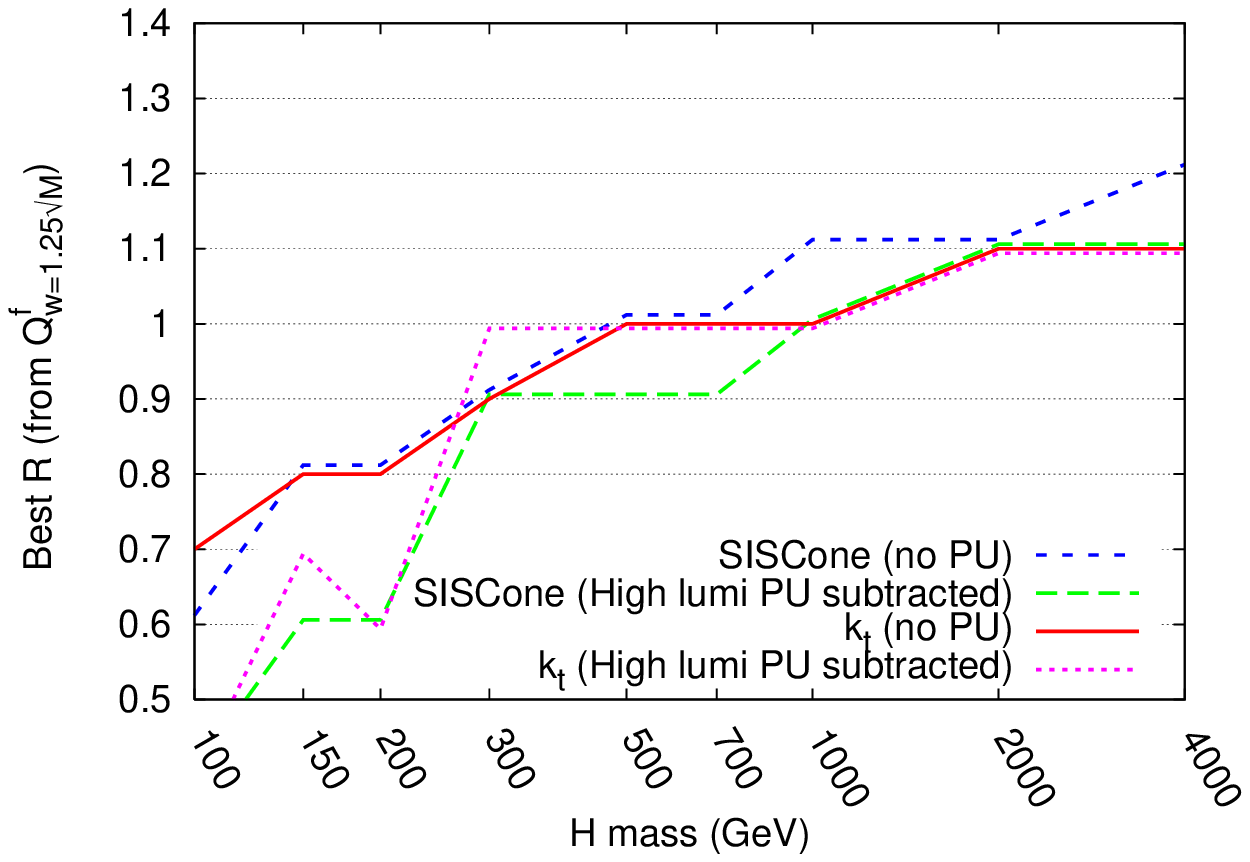}
\caption{
Left plot: the best value of the jet radius $R_{\rm best}$ 
as determined for
gluon jets  as a function of the relevant
mass scale. Right plot: 
comparison of the optimal $R_{\rm best}$ for the SISCone and
$k_T$ algorithms for gluon jets in without PU case and in the high-lumi PU
case with subtraction \label{fig:RbestPU}}
\end{figure}
\end{center}

 

\begin{footnotesize}


\end{footnotesize}


\end{document}